\documentclass[onecollarge]{svjour2}       
\smartqed  
\usepackage{graphicx}

\usepackage{amsmath,amssymb}
\journalname{Journal of Statistical Physics}
\begin{document}

\title{How does pressure fluctuate in equilibrium? 
}


\author{Ken Hiura         \and
        Shin-ichi Sasa 
}


\institute{K. Hiura \and S. Sasa \at
              Department of Physics, Kyoto University, Kyoto 606-8502, Japan \\
              \email{hiura.ken.88n@st.kyoto-u.ac.jp, sasa@scphys.kyoto-u.ac.jp}           
}

\date{Received: date / Accepted: date}

\maketitle

\begin{abstract}
We study fluctuations of pressure in equilibrium for classical particle systems. In equilibrium statistical mechanics, pressure for a microscopic state is defined by the derivative of a thermodynamic function or, more mechanically, through the momentum current. We show that although the two expectation values converge to the same equilibrium value in the thermodynamic limit, the variance of the mechanical pressure is in general greater than that of the pressure defined through the thermodynamic relation. We also present a condition for experimentally detecting the difference between them in an idealized measurement of momentum transfer.
\end{abstract}

\section{Introduction}
\label{intro}
Thermodynamic fluctuation theory plays important roles in the theory of critical phenomena and linear response theory. The second-order phase transition is characterized by divergence of the correlation length of the order parameter and the linear transport coefficient is determined by the time correlation function of the corresponding current in equilibrium. The general theory of thermodynamic fluctuation in equilibrium seems to be well established. However, there is a subtle problem with static fluctuations of intensive variables in equilibrium as described below. In this paper, we investigate this subject by analyzing the pressure fluctuation.

The static fluctuation theory in equilibrium was initiated by Einstein \cite{einstein} and is based on the Boltzmann formula and the principle of equal probability. Given a set of macroscopic variables $X$ characterizing an isolated system under internal constraints, the number (the phase space volume for classical particle systems) of microstates corresponding to a macrostate $X=x$, $W(x)$, is related to the entropy of the system for the macrostate $x$, $S(x)$, by the Boltzmann formula; i.e., $W(x) \propto \mathrm{e}^{S(x)/k_{\mathrm{B}}}$. According to the principle of equal probability, all the microstates consistent with $x$ are equally probable; i.e., $\mathrm{Prob}(X=x) \propto W(x)$. The entropy function therefore uniquely determines the joint probability distribution of $X$, and the equilibrium state under the given constraints is characterized by the most probable value of $X$. Similarly, the probability distribution of a set of macroscopic variables of a subsystem that is described by a statistical ensemble (e.g., the canonical ensemble and the grand canonical ensemble) is also determined by the corresponding thermodynamic function \cite{callen}.

All random variables in the above formulation are mechanical; i.e., functions of the microscopic states. From the viewpoint of statistical mechanics, the fluctuation of a mechanical variable is conceptually clear because the equilibrium state is described by a probability distribution on microstates. Nonetheless, Landau and Lifshitz calculated fluctuations of nonmechanical thermodynamic variables, such as entropy and temperature \cite{landau}. These variables are not functions of the microstates. In fact, to calculate variances of nonmechanical variables, they defined ``fluctuating entropy/temperature/pressure'' with the aid of thermodynamic relations. Although this theory has been accepted, the relation to the fluctuation measured in experiments remains to be studied. The motivation of the present paper is to investigate the validity of Landau and Lifshitz's result in experimental situations.

This paper focuses on pressure fluctuations because the pressure is mechanically defined through momentum conservation. In pressure measurements, without using thermodynamic relations, we directly observe the momentum flux flowing into the measurement device from the system. In molecular dynamics simulations, the mechanical pressure is commonly measured as the pressure of the system. It is thus not obvious whether Landau and Lifshitz's argument leads to a good prediction of the pressure fluctuation in experiments.

The remainder of the paper is organized as follow. In section \ref{sec:LL}, we reformulate the argument for calculating the fluctuation of temperature and pressure introduced by Landau and Lifshitz. Our main result is presented in sections \ref{sec:pressure} and \ref{sec:exp}. We state and prove (in)equalities connecting the fluctuations of pressure defined through the thermodynamic relation and that of the mechanical pressure in section \ref{sec:pressure}. We also present a condition for experimentally detecting the difference in section {\ref{sec:exp}.

\section{Landau and Lifshitz's argument}
\label{sec:LL}
This section reformulates the argument made by Landau and Lifshitz \cite{landau}. For a given thermodynamic system (e.g., a single-component simple fluid system), the entropy function $S(E,V,N)$ of the system is determined experimentally, where $E$ is the total internal energy, $V$ the total volume and $N$ the total number of particles of the system. Temperature and pressure functions are defined as derivatives of the thermodynamic function: $T=(\partial S/\partial E)_{V,N}^{-1}$ and $P=T(\partial S/\partial V)_{E,N}$.

We take a subsystem that is much smaller than the overall system but still contains a sufficient number of particles so that the surface effect is negligible. A microstate of the subsystem, consisting of $N$ particles in a region $\Lambda$ with volume $V$ in $d$-dimensional space, is specified by a point $\Gamma =(q_1, \dots, q_N, p_1, \dots, p_N)$, where $q_i \in \Lambda$ and $p_i \in \mathbb{R}^d$ are respectively coordinates and momenta of the $i$-th particles.  When the number of particles $N$ in the subsystem is fixed and the volume $V$ fluctuates, the macrostate of the subsystem is identified with the T-p ensemble; i.e., the constant-temperature and constant-pressure ensemble. More precisely, the system is described by a probability distribution specified by the temperature $T$, pressure $P$, and number of particles $N$ in the subsystem:
\begin{equation}
 \mu^{\mathrm{tp}}_{T,P,N}(\Gamma, \hat{V}) \propto \exp \left( - \frac{H(\Gamma) + P \hat{V}}{k_{\mathrm{B}}T} \right),
\end{equation}
where $H(\Gamma)$ and $\hat{V}$ are respectively the Hamiltonian and the total volume of the subsystem. We note that $\hat{V}$ is a random variable. The expectation value of a mechanical variable $A(\Gamma, \hat{V})$ with respect to the T-p ensemble is denoted by $\left\langle A \right\rangle^{\mathrm{tp}}_{T,P,N}$. In particular, we obtain the joint probability distribution of the energy and volume, $\mathbb{P}^{\mathrm{tp}}_{T,P,N}(\mathcal{E},\mathcal{V})$, from the T-p ensemble as
\begin{align}
 \mathbb{P}^{\mathrm{tp}}_{T,P,N}(\mathcal{E},\mathcal{V}) & =  \left\langle \delta (\mathcal{E}-H) \delta(\mathcal{V}-\hat{V}) \right\rangle^{\mathrm{tp}}_{T,P,N}
 \nonumber \\
\label{eq:ltp}
 & \propto \exp \left( - \frac{\Delta \mathcal{E} + P \Delta \mathcal{V} - T \Delta S}{k_{\mathrm{B}}T} \right),
\end{align}
where $\Delta \mathcal{E} = \mathcal{E} - \left\langle H \right\rangle^{\mathrm{tp}}_{T,P,N}$, $\Delta \mathcal{V} = \mathcal{V} - \langle \hat{V} \rangle^{\mathrm{tp}}_{T,P,N}$ and $\Delta S = S(\mathcal{E},\mathcal{V},N) - S( \left\langle H \right\rangle^{\mathrm{tp}}_{T,P,N}, \langle \hat{V} \rangle^{\mathrm{tp}}_{T,P,N}, N)$ are displacements from equilibrium values. The formula (\ref{eq:ltp}) is identical to (112.2) in \cite{landau}.

However, thermodynamic variables, such as entropy and temperature, are not mechanical variables because these concepts are not formulated in mechanics. Landau and Lifshitz introduced random variables corresponding to entropy, temperature, and pressure as functions of original (mechanical) random variables, $\mathcal{E}$ and $\mathcal{V}$, with the aid of thermodynamic relations. More explicitly, they defined fluctuating entropy, temperature, and pressure by
\begin{align}
 \label{eq:le}
 \mathcal{S}(\mathcal{E}, \mathcal{V}) &= S(\mathcal{E},\mathcal{V},N),
 \\
 \label{eq:lt}
 \mathcal{T}(\mathcal{E},\mathcal{V}) &= \left( \frac{\partial S(\mathcal{E},\mathcal{V},N)}{\partial \mathcal{E}} \right)^{-1},
 \\
 \label{eq:lp}
 \mathcal{P}(\mathcal{E},\mathcal{V}) &= \mathcal{T}(\mathcal{E},\mathcal{V})  \frac{\partial S(\mathcal{E},\mathcal{V},N)}{\partial \mathcal{V}}.
\end{align}
We note that the random variables, $\mathcal{T}$ and $\mathcal{P}$, are objects mathematically different from $T$ and $P$, the parameters specifying the T-p ensemble. The probability distribution of $\mathcal{S}$, $\mathcal{T}$, and $\mathcal{P}$ are obtained from that of original variables (\ref{eq:ltp}). In particular, within the Gaussian approximation, two random variables appropriately chosen follow a normal distribution (e.g. $\mathcal{T}$, and $\mathcal{V}$ follow (112.4) in \cite{landau}). Landau and Lifshitz calculated the variances of $\mathcal{S}$, $\mathcal{T}$, and $\mathcal{P}$ on the basis that any linear transformation preserves  normality. For example, the variance of $\mathcal{P}$ is given by
\begin{align}
 \label{eq:thpflu}
 \left\langle \left(\mathcal{P} - \left\langle \mathcal{P} \right\rangle^{\mathrm{tp}}_{T,P,N} \right)^2 \right\rangle^{\mathrm{tp}}_{T,P,N} = - k_{\mathrm{B}}T \left( \frac{\partial P}{\partial V} \right)_S + o(N^{-1}).
\end{align}
The probability distribution of $\mathcal{S}$, $\mathcal{T}$, and $\mathcal{P}$ depends on that of original random variables. When the system is described by the canonical ensemble $\mu^{\mathrm{can}}_{T,V,N}(\Gamma,\hat{V}) \propto \mathrm{e}^{- H(\Gamma) / k_{\mathrm{B}}T} \delta(V-\hat{V})$, specified by the temperature $T$, volume $V$, and number of particles $N$, the variance of $\mathcal{P}$ is given by
\begin{equation}
\label{eq:thpcanflu}
 \left\langle \left(\mathcal{P} - \left\langle \mathcal{P} \right\rangle^{\mathrm{can}}_{T,V,N} \right)^2 \right\rangle^{\mathrm{can}}_{T,V,N} = \frac{k_{\mathrm{B}}T^2}{C_V} \left( \frac{\partial P}{\partial T} \right)_V^2 + o(N^{-1}),
\end{equation}
where $\left\langle A \right\rangle^{\mathrm{can}}_{T,V,N}$ is the expectation value of $A(\Gamma,\hat{V})$ with respect to $\mu^{\mathrm{can}}_{T,V,N}$, and $C_V$ is the heat capacity at constant volume. This theory gives a systematic procedure for calculating fluctuations of intensive variables defined as (\ref{eq:le}), (\ref{eq:lt}) and (\ref{eq:lp}) in each ensemble. A complete set of variances of thermodynamic variables in each ensemble is presented by Ref. \cite{mishin}.

There is an alternative definition of a random variable corresponding to an intensive variable (where we restrict ourselves to pressure). We introduce the mechanical variable
\begin{align}
 \label{eq:mcp}
 \tilde{P}(\Gamma,\hat{V}) = \left( \frac{\partial S(E,V,N)}{\partial E} \right)^{-1} \left. \frac{\partial S(E,V,N)}{\partial V} \right|_{E=H(\Gamma),V=\hat{V}},
\end{align}
which is regarded as pressure for a microstate $\Gamma$ and volume $\hat{V}$. As we shall show later, (\ref{eq:mcp}) is identified with the microcanonical expectation value of the mechanical pressure whose definition will be given in (\ref{eq:mp}). We therefore refer to it as \textit{microcanonical pressure} in this paper. The definition (\ref{eq:mcp}) is equivalent to (\ref{eq:lp}) in the sense that both random variables obey the same distribution. In particular, the variance of $\tilde{P}$ in the T-p ensemble is given by
\begin{align}
 \label{eq:mcpfluflu}
 \left\langle \left(\tilde{P} - \langle \tilde{P} \rangle^{\mathrm{tp}}_{T,P,N} \right)^2 \right\rangle^{\mathrm{tp}}_{T,P,N} = - k_{\mathrm{B}}T \left( \frac{\partial P}{\partial V} \right)_S + o(N^{-1}),
\end{align}
which is equivalent to (\ref{eq:thpflu}). Although it seems that there is no need to introduce (\ref{eq:mcp}) separately from (\ref{eq:lp}), the mechanical definition (\ref{eq:mcp}) is of importance in clarifying what is implicitly assumed at the level of microscopic mechanics in definition (\ref{eq:lp}). A crucial point is that the microcanonical pressure is constant on any constant-energy surface, $\Sigma(E) = \{ \Gamma \mid H(\Gamma)=E \}$, which implies that we cannot describe fluctuations on $\Sigma(E)$ using the definitions (\ref{eq:lp}) and (\ref{eq:mcp}). In fact, the microcanonical pressure and mechanical pressure are qualitatively different in this respect. See (\ref{eq:canflu}) and (\ref{eq:tpflu}).

This method for calculating fluctuations of intensive variables has been used, for instance, in mode-coupling theory \cite{ernst,kawasaki}. However, several authors have objected to the calculation method \cite{munster,wallace} or the concept itself of fluctuations of intensive variables \cite{kittel} for many years. M{\"u}nster \cite{munster} reported that for a monatomic ideal gas, the variance of the pressure obtained by Landau and Lifshitz (\ref{eq:thpflu}) is different from that of the mechanical pressure calculated using the virial theorem and the fluctuation formula of the energy in the canonical ensemble. However, as we shall show later, the two variances in the same ensemble are identical for a monatomic ideal gas. The difference reported simply arises from the nonequivalence of ensemble for variances. Wallace \cite{wallace} derived ensemble transformation formulae for the mechanical pressure and concluded that the result (\ref{eq:thpflu}) differs from the variance of the mechanical pressure for general cases. Although the observation is indeed correct, the general relationship between the fluctuations of mechanical pressure and microcanonical pressure has never been presented, but will be revealed in section \ref{sec:pressure} through the formulation of the theory of Landau and Lifshitz given in this section.

A more important problem is whether the variances of pressure obtained in experiments are identical to (\ref{eq:thpflu}) and (\ref{eq:mcpfluflu}). We discuss the problem in section \ref{sec:exp}.

\section{Mechanical pressure versus microcanonical pressure}
\label{sec:pressure}
\subsection{Main result}
An important observation is that pressure is defined mechanically by the momentum current density. The explicit functional form of the momentum current density $j^{ab}(q,\Gamma)$ is determined by the continuity equation of the momentum density $g^a(q,\Gamma)=\sum_{i=1}^N p_i \delta(q-q_i)$. In homogeneous equilibrium systems, the stress tensor $\tau^{ab}(\Gamma)$ is defined by the space average of $j^{ab}(x;\Gamma)$ over the total region and the pressure given by $P_m(\Gamma) = (1/d) \sum_{a=1}^d \tau^{aa}(\Gamma)$. For a classical particle system with short-range pairwise interaction $\varphi(|q|)$, the pressure is given by
\begin{equation}
\label{eq:mp}
 P_m(\Gamma,\hat{V}) = \frac{1}{d \hat{V}} \left( \sum_{i=1}^N \frac{p_i^2}{m} - \frac{1}{2} \sum_{i \neq j} (q_i - q_j) \frac{\partial}{\partial q_i}\varphi (|q_i-q_j|) \right).
\end{equation}
We refer to it as \textit{mechanical pressure} to distinguish it from the microcanonical pressure.

The main result of the paper is
\begin{equation}
\label{eq:main}
  \left\langle \left( P_m - \left\langle P_m \right\rangle^{\mathrm{tp}}_{T,P,N} \right)^2  \right\rangle^{\mathrm{tp}}_{T,P,N} \geq \left\langle \left( \tilde{P} - \langle \tilde{P} \rangle^{\mathrm{tp}}_{T,P,N} \right)^2  \right\rangle^{\mathrm{tp}}_{T,P,N}+ o(N^{-1}),
\end{equation}
although $\left\langle P_m \right\rangle^{\mathrm{tp}}_{T,P,N}=\langle \tilde{P} \rangle^{\mathrm{tp}}_{T,P,N} + O(N^{-1})$. In particular, for general cases, the fluctuation of the mechanical pressure is strictly greater than that of the microcanonical pressure.

\subsection{Expectation value of pressure: virial theorem}
We first review basic properties of the expectation values of pressure. The virial theorem states that the statistical mechanical expectation value of mechanical pressure coincides with derivatives of the thermodynamic function. For the canonical ensemble, the virial theorem states
\begin{equation}
\label{eq:virial}
 \left\langle P_m \right\rangle^{\mathrm{can}}_{T,V,N} = - \frac{\partial F(T,V,N)}{\partial V},
\end{equation}
where $F(T,V,N)$ is the free energy function that is defined from the partition function of the system. Similarly, the virial theorem for the microcanonical ensemble states
\begin{equation}
\label{eq:mcvirial}
 \left\langle P_m \right\rangle^{\mathrm{mc}}_{E,V,N} = \left( \frac{\partial S(E,V,N)}{\partial E} \right)^{-1} \frac{\partial S(E,V,N)}{\partial V} + O(N^{-1}),
\end{equation}
where $\left\langle A \right\rangle^{\mathrm{mc}}_{E,V,N}$ is the expectation value of $A(\Gamma,\hat{V})$ with respect to the microcanonical ensemble, $\mu^{\mathrm{mc}}_{E,V,N} \propto \delta(E-H(\Gamma)) \delta(V-\hat{V})$, specified by the energy $E$, volume $V$, and number of particles $N$, and $S(E,V,N)$ is the entropy function obtained from the Boltzmann formula. We repeat that the microcanonical pressure for a microstate $\Gamma$ is identical to the expectation value of the mechanical pressure with respect to $\mu^{\mathrm{mc}}_{H(\Gamma),\hat{V},N}$:
\begin{equation}
\label{eq:mcmc}
 \tilde{P}(\Gamma, \hat{V}) = \left\langle P_m \right\rangle^{\mathrm{mc}}_{H(\Gamma),\hat{V},N} + O(N^{-1}).
\end{equation}
In general, however, the microcanonical pressure $\tilde{P}(\Gamma,\hat{V})$ is different from the mechanical pressure $P_m(\Gamma,\hat{V})$.

We get from (\ref{eq:mcmc}) that
\begin{align}
\label{eq:meqth}
 \left\langle P_m \right\rangle^{\mathrm{can}}_{T,V,N} = \langle \tilde{P} \rangle^{\mathrm{can}}_{T,V,N} +O(N^{-1})
\end{align}
and
\begin{equation}
\label{eq:ppp}
  \left\langle P_m \right\rangle^{\mathrm{tp}}_{T,P,N} = \langle \tilde{P} \rangle^{\mathrm{tp}}_{T,P,N} + O(N^{-1}).
\end{equation}
Consequently, with respect to the expectation value, both the mechanical pressure and microcanonical pressure give the same value in the thermodynamic limit.

We assume the equivalence of ensembles for $P_m$ and $\tilde{P}$ between the microcanonical and canonical ensemble and between the canonical and T-p ensemble: for $Q=P_m$ and $\tilde{P}$,
\begin{equation}
 \left\langle Q \right\rangle^{\mathrm{can}}_{T,V,N} = \left\langle Q \right\rangle^{\mathrm{mc}}_{E(T,V,N),V,N}+O(N^{-1}),
\end{equation}
and
\begin{align}
 \left\langle Q \right\rangle^{\mathrm{tp}}_{T,P,N} &= \left\langle Q \right\rangle^{\mathrm{can}}_{T,V(T,P,N),N}+O(N^{-1})
 \\
 & = \left\langle Q \right\rangle^{\mathrm{mc}}_{E(T,P,N),V(T,P,N),N}+O(N^{-1}),
\end{align}
where $E(T,V,N) = \langle H \rangle^{\mathrm{can}}_{T,V,N}$, $V(T,P,N) = \langle \hat{V} \rangle^{\mathrm{tp}}_{T,P,N}$ and $E(T,P,N)=E(T,V(T,P,N),N)$. 

\subsection{Variance of pressure}
\label{subsec:vp}
We present the main results. In contrast to the expectation value, the variance of the mechanical pressure is different from that of the microcanonical pressure in general. In fact, we can show that
\begin{align}
\label{eq:canflu}
 \left\langle  \left( P_m - \left\langle P_m \right\rangle^{\mathrm{can}}_{T,V,N} \right)^2  \right\rangle^{\mathrm{can}}_{T,V,N} = & \left\langle  \left( \tilde{P} - \langle \tilde{P} \rangle^{\mathrm{can}}_{T,V,N} \right)^2 \right\rangle^{\mathrm{can}}_{T,V,N} 
 \notag \\
 &+ \left\langle  \left( P_m - \left\langle P_m \right\rangle^{\mathrm{mc}}_{E(T,V,N),V,N} \right)^2 \right\rangle^{\mathrm{mc}}_{ E(T,V,N),V,N} + o(N^{-1}),
\end{align}
and
\begin{align}
\label{eq:tpflu}
 \left\langle  \left( P_m - \left\langle P_m \right\rangle^{\mathrm{tp}}_{T,P,N} \right)^2  \right\rangle^{\mathrm{tp}}_{T,P,N} & =  \left\langle  \left( \tilde{P} - \langle \tilde{P} \rangle^{\mathrm{tp}}_{T,P,N} \right)^2 \right\rangle^{\mathrm{tp}}_{T,P,N} 
 \notag \\ 
 & + \left\langle  \left( P_m - \left\langle P_m \right\rangle^{\mathrm{mc}}_{E(T,P,N),V(T,P,N),N} \right)^2 \right\rangle^{\mathrm{mc}}_{ E(T,P,N),V(T,P,N),N} + o(N^{-1}).
\end{align}
Equality (\ref{eq:tpflu}) leads to inequality (\ref{eq:main}). We note that all the terms in (\ref{eq:canflu}) and (\ref{eq:tpflu}) are of order $O(N^{-1})$ and in particular the second term on the right-hand side are of the same order as the first term, the fluctuation obtained by Landau and Lifshitz. This type of relation also holds for the grand canonical ensemble.

For ideal gases the mechanical pressure has no fluctuation in the microcanonical ensemble because $P_m(\Gamma,\hat{V})$ is proportional to the Hamiltonian $H(\Gamma)$ for any $\Gamma$ and fixed $\hat{V}$. In this case, the fluctuations of the mechanical and microcanonical pressures are identical in any ensemble. Except for these special cases, the second term on the right-hand side of (\ref{eq:tpflu}) gives the extra contribution beyond the fluctuation of the microcanonical pressure. Therefore, the fluctuation of the mechanical pressure is strictly greater than that of the microcanonical pressure in the canonical and T-p ensembles. These inequalities are easily understood from the fact that the mechanical pressure fluctuates on the constant-energy surface unlike the microcanonical pressure.

We give a proof of the relations (\ref{eq:canflu}) and (\ref{eq:tpflu}). We use a general equality. For any mechanical variable $A(\Gamma,\hat{V})$, the relation between the variances
\begin{align}
\label{eq:gibbs}
  \left\langle \left( A - \left\langle A \right\rangle^{\mathrm{can}}_{T,V,N} \right)^2 \right\rangle^{\mathrm{can}}_{T,V,N} =& \left\langle \left( A - \left\langle A \right\rangle^{\mathrm{mc}}_{H,V,N} \right)^2 \right\rangle^{\mathrm{can}}_{T,V,N} + \left\langle \left( \left\langle A \right\rangle^{\mathrm{mc}}_{H,V,N} - \left\langle A \right\rangle^{\mathrm{can}}_{T,V,N} \right)^2 \right\rangle^{\mathrm{can}}_{T,V,N}
\end{align}
holds \cite{gibbs}. The first term on the right-hand side is the expectation value of the fluctuation on the constant-energy surface with respect to the canonical ensemble, while the second term is the variance of the quantity averaged over the constant-energy surface. By using
\begin{align}
 & \left\langle \left( A - \left\langle A \right\rangle^{\mathrm{mc}}_{H,V,N} \right)\left( \left\langle A \right\rangle^{\mathrm{mc}}_{H,V,N} - \left\langle A \right\rangle^{\mathrm{can}}_{T,V,N} \right) \right\rangle^{\mathrm{can}}_{T,V,N} 
 \notag \\
 &= \left\langle \left\langle \left( A - \left\langle A \right\rangle^{\mathrm{mc}}_{H,V,N} \right)\left( \left\langle A \right\rangle^{\mathrm{mc}}_{H,V,N} - \left\langle A \right\rangle^{\mathrm{can}}_{T,V,N} \right) \right\rangle^{\mathrm{mc}}_{H,V,N} \right\rangle^{\mathrm{can}}_{T,V,N}
 \notag \\
 &= \left\langle \left\langle \left( A - \left\langle A \right\rangle^{\mathrm{mc}}_{H,V,N} \right) \right\rangle^{\mathrm{mc}}_{H,V,N} \left( \left\langle A \right\rangle^{\mathrm{mc}}_{H,V,N} - \left\langle A \right\rangle^{\mathrm{can}}_{T,V,N} \right)  \right\rangle^{\mathrm{can}}_{T,V,N}
 \notag \\
 &=0,
\end{align}
we obtain (\ref{eq:gibbs}). The equality (\ref{eq:gibbs}) is easily extended to the T-p ensemble and the grand canonical ensemble. For instance, the equality
\begin{align}
\label{eq:gibbs2}
  \left\langle \left( A - \left\langle A \right\rangle^{\mathrm{tp}}_{T,P,N} \right)^2 \right\rangle^{\mathrm{tp}}_{T,P,N} =& \left\langle \left( A - \left\langle A \right\rangle^{\mathrm{mc}}_{H,V,N} \right)^2 \right\rangle^{\mathrm{tp}}_{T,P,N} + \left\langle \left( \left\langle A \right\rangle^{\mathrm{mc}}_{H,V,N} - \left\langle A \right\rangle^{\mathrm{tp}}_{T,P,N} \right)^2 \right\rangle^{\mathrm{tp}}_{T,P,N}
\end{align}
holds.

By setting $A=P_m$ in (\ref{eq:gibbs}) and by using (\ref{eq:mcmc}) and (\ref{eq:meqth}), we obtain
\begin{align}
\label{eq:prf1}
  \left\langle \left( P_m - \left\langle P_m \right\rangle^{\mathrm{can}}_{T,V,N} \right)^2 \right\rangle^{\mathrm{can}}_{T,V,N} = & \left\langle \left( P_m - \left\langle P_m \right\rangle^{\mathrm{mc}}_{H,V,N} \right)^2 \right\rangle^{\mathrm{can}}_{T,V,N}
\notag \\
&+ \left\langle \left( \tilde{P} - \langle \tilde{P} \rangle^{\mathrm{can}}_{T,V,N} \right)^2 \right\rangle^{\mathrm{can}}_{T,V,N} + o(N^{-1}),
\end{align}
where we have used the estimation
\begin{align}
 \left\langle \left( \tilde{P} - \langle \tilde{P} \rangle^{\mathrm{can}}_{T,V,N} +O(N^{-1}) \right)^2 \right\rangle^{\mathrm{can}}_{T,V,N} = \left\langle \left( \tilde{P} - \langle \tilde{P} \rangle^{\mathrm{can}}_{T,V,N} \right)^2 \right\rangle^{\mathrm{can}}_{T,V,N} + o(N^{-1}).
\end{align}
Furthermore, by employing the saddle point method, we get
\begin{align}
\label{eq:prf2}
 \left\langle \left( P_m - \left\langle P_m \right\rangle^{\mathrm{mc}}_{H,V,N} \right)^2 \right\rangle^{\mathrm{can}}_{T,V,N} = \left\langle \left( P_m - \left\langle P_m \right\rangle^{\mathrm{mc}}_{E(T,V,N),V,N} \right)^2 \right\rangle^{\mathrm{mc}}_{E(T,V,N),V,N} + o(N^{-1}).
\end{align}
We then obtain (\ref{eq:canflu}) from (\ref{eq:prf1}) and (\ref{eq:prf2}). Similarly, we get (\ref{eq:tpflu}) from (\ref{eq:mcmc}), (\ref{eq:ppp}) and (\ref{eq:gibbs2}).

\section{Pressure fluctuation in experiment}
\label{sec:exp}
We consider the fluctuation of the pressure measured in experiments. In particular, we investigate whether the prescription proposed by Landau and Lifshitz works well in experiments. As a typical method of measuring pressure, elastic materials may be used as pressure transducers. We measure the strain of the material induced by the pressure of the system and obtain the pressure value from the strain with the aid of a known elastic property of the material; e.g., elastic modulus. What is measured in this experimental setup is the momentum transfer into the measurement device from the system. We should therefore use the mechanical pressure to analyze the measurement data. Moreover, the measurement process is not instantaneous. Because the response time of transducers is finite, it is impossible to measure the pressure during a time interval less than the response time.

We idealize this situation by introducing a small rigid sphere $\Omega$ as a probe. The particles of the system collide elastically at the boundary of the sphere $\partial \Omega$ and we can monitor the momentum transfer into the sphere $\Omega$ from particles due to elastic collisions at the boundary. In other words, the measurement value of pressure in this experiment is given by the momentum transfer per unit time and unit area into the sphere $\Omega$ from the system in a time interval $\tau$, which is denoted by $\mathcal{G}^{\tau}$. Explicitly, we let $( q_i(t), p_i(t) )_{i=1}^{N}$ be the microstate of the system at time $t$ that is the solution of the Hamilton equations with elastic collisions at $\partial \Omega$ for an initial state $\Gamma$ at $t=0$ and $\{( i_j, t_j )\}_{j=1}^{M}$ be a series of pairs of the index of the collision particle and collision time in the time interval $[0,\tau]$. $\mathcal{G}^{\tau}(\Gamma)$ is then expressed as
\begin{align}
\label{eq:mt}
 \mathcal{G}^{\tau}(\Gamma) = \frac{1}{\tau} \frac{1}{|\partial \Omega|} \sum_{j=1}^{M} 2 (p_{i_j}(t_j) \cdot \hat{\omega}(q_{i_j}(t_j))),
\end{align}
where $\hat{\omega}(q)$ is the unit outward normal vector at $q \in \partial \Omega$.

We here specify the measurement scheme as follow. We prepare a simple fluid system with the T-p ensemble $\mu^{\mathrm{tp}}_{T,P,N}$ and set the probe $\Omega$ in the system. We sample an initial state $\Gamma$ according to the T-p ensemble $\mu^{\mathrm{tp}}_{T,P,N}$ and measure the momentum transfer $\mathcal{G}^{\tau}(\Gamma)$ into the probe $\Omega$ over the time interval $\tau$. Repeating this procedure many times, we obtain the empirical distribution of (\ref{eq:mt}). We can then calculate the variance of $\mathcal{G}^{\tau}$ from the distribution.

We assume the two properties of (\ref{eq:mt}). The first assumption is
\begin{align}
\label{eq:mte}
 \left\langle \mathcal{G}^{\tau} \right\rangle = \left\langle P_m \right\rangle,
\end{align}
where $\langle A \rangle$ is the expectation value of $A(\Gamma)$ with respect to equilibrium ensembles. It should be noted that (\ref{eq:mte}) holds for any $\tau > 0$. A rigorous treatment of assumption (\ref{eq:mte}) for the case of momentum transfer into the bound wall was presented by Ref. \cite{presutti}. The second assumption is that the long-time limit of (\ref{eq:mt}) is asymptotically equal to the expectation value of the mechanical pressure almost surely with respect to the microcanonical ensemble; i.e., we assume that
\begin{align}
\label{eq:mtt}
 \mathcal{G}^{\tau} (\Gamma) \to \left\langle P_m \right\rangle^{\mathrm{mc}}_{E,V,N} \quad  \text{as} \quad  \tau \to \infty
\end{align}
for almost all initial states $\Gamma$ with respect to the microcanonical ensemble $\mu^{\mathrm{mc}}_{E,V,N}$. These assumptions ensure that we get the equilibrium value for pressure in this measurement from the ensemble average and the long-time average. 

Under the assumption (\ref{eq:mte}), by repeating a procedure similar to that argued in section \ref{subsec:vp}, we obtain
\begin{align}
\label{eq:dg}
  \left\langle \left( \mathcal{G}^{\tau} - \left\langle \mathcal{G}^{\tau} \right\rangle^{\mathrm{tp}}_{T,P,N} \right)^2 \right\rangle^{\mathrm{tp}}_{T,P,N} =& \left\langle  \left( \tilde{P} - \langle \tilde{P} \rangle^{\mathrm{tp}}_{T,P,N} \right)^2 \right\rangle^{\mathrm{tp}}_{T,P,N} 
 \notag \\
  & + \left\langle \left( \mathcal{G}^{\tau} - \left\langle \mathcal{G}^{\tau} \right\rangle^{\mathrm{mc}}_{H,V,N} \right)^2 \right\rangle^{\mathrm{tp}}_{T,P,N} + o(N^{-1}).
\end{align}
Assumption (\ref{eq:mtt}) implies that the second term on the right-hand side decreases as the measurement time $\tau$ increases; i.e.,  
\begin{align}
 \lim_{\tau \to \infty} \left\langle  \left( \mathcal{G}^{\tau} - \left\langle \mathcal{G}^{\tau} \right\rangle^{\mathrm{mc}}_{H,V,N} \right)^2 \right\rangle^{\mathrm{tp}}_{T,P,N} = \lim_{\tau \to \infty} \left\langle \left\langle  \left( \mathcal{G}^{\tau} - \left\langle \mathcal{G}^{\tau} \right\rangle^{\mathrm{mc}}_{H,V,N} \right)^2 \right\rangle^{\mathrm{mc}}_{H,V,N} \right\rangle^{\mathrm{tp}}_{T,P,N} = 0.
\end{align}
Therefore, the fluctuation of the measurement value asymptotically approaches the result of Landau and Lifshitz in the long measurement time limit.

Which term of the two on the right-hand side of (\ref{eq:dg}) gives a dominant contribution to the fluctuation measured in experiments in finite measurement time depends on $\tau$. We let $R$ and $L = (\left\langle V \right\rangle^{\mathrm{tp}}_{T,P,N})^{1/d}$ be the linear dimensions of the probe and the total system. We consider the macroscopic probe: that is, we observe the asymptotic behavior of the fluctuation in the thermodynamic limit $R \to \infty$ and $L \to \infty$ with the ratio $R/L$ fixed. We also assume that the ratio $R/L$ is sufficiently small such that the boundary effect of the total system on the fluctuation at the surface of the probe is negligibly small, and consider the three-dimensional case $d=3$. For dilute gases, the central limit theorem naively implies that
\begin{align}
 \left\langle \left( \mathcal{G}^{\tau} - \left\langle \mathcal{G}^{\tau} \right\rangle \right)^2 \right\rangle \propto \frac{1}{\tau R^2}.
\end{align}
In contrast, for dense fluids, the situation is completely different. In this case, the central limit theorem with respect to the area average is violated due to the correlation between the particles. In fact, the fluctuation at the surface is suppressed and
\begin{align}
\label{eq:hyper}
 \left\langle \left( \mathcal{G}^{\tau} - \left\langle \mathcal{G}^{\tau} \right\rangle \right)^2 \right\rangle = \frac{c k_{\mathrm{B}}T \eta}{\tau R^3}
\end{align}
to leading order \cite{itami}, where $\eta$ is the shear viscosity and $c$ is a numerical constant ( $c=3/4\pi$ for stick boundary condition). We do not specify the ensemble in (\ref{eq:hyper}) since the choice of ensembles does not affect the following argument. We note that the formula (\ref{eq:hyper}) is valid for $\tau$ such that $\tau_\mathrm{m} \ll \tau \ll \tau_\mathrm{M}$, where $\tau_{\mathrm{m}}$ is the correlation time of the pressure and $\tau_{\mathrm{M}}$ is the relaxation time of the momentum density. The ratio of (\ref{eq:thpflu}) to (\ref{eq:hyper}), which determines which term on the right-hand side of (\ref{eq:dg}) is dominant, is $r = \tau l /  \eta \kappa_s$, where $l=c(R/L)^3$ and $\kappa_s$ is the adiabatic compressibility. For a given set of material and probe, the ratio $r$ is determined by the measurement time $\tau$. When $r \ll 1$, the second term is dominant and the contribution from the bulk fluctuation is invisible in experiments. In other words, the prescription proposed by Landau and Lifshitz does not give a good prediction of experimental results. When $r \sim 1$, two contributions are comparable. We define $\tau_c$ by the measurement time such that $r=1$. For water at room temperature ($\kappa_s = 4.5 \times 10^{-9} \ \mathrm{Pa}^{-1}$ and $\eta = 1.0 \times 10^{-3} \ \mathrm{Pa \cdot s}$) and $l=10^{-12}$, $\tau_c = 0.45 \ \mathrm{s}$. Strictly speaking, this time scale is beyond the range of application of the formula (\ref{eq:hyper}), and a more detailed analysis of the fluctuation is needed to evaluate precisely the measurement time such that the two contributions are comparable. We believe, however, that it would give a reasonable estimation of the measurement time necessary to observe experimentally the fluctuation of pressure that is incompatible with the result of Landau and Lifshitz.

\section{Conclusion}
We have studied fluctuations of pressure in equilibrium. We have reformulated the conventional thermodynamic fluctuation theory of nonmechanical variables developed by Landau and Lifshitz within equilibrium statistical mechanics. In this theory, pressure is defined through the thermodynamic relation; that is, the pressure is the expectation value of the mechanical pressure, which is defined by using the momentum current, with respect to the microcanonical ensemble. We have refered to it as microcanonical pressure.

We have shown that although the expectation values in equilibrium of the mircocanonical pressure and the mechanical pressure are identical, the fluctuation of the mechanical pressure in the T-p ensemble contains the fluctuation in the microcanonical ensemble in addition to the fluctuation of the microcanical pressure obtained by Landau and Lifshitz. Since the instantaneous value of the mechanical pressure is calculated in molecular dynamics simulations, it is possible to evaluate the difference between the fluctuations of the mechanical pressure and the microcanonical pressure.

Meanwhile, whether they can be distinguished in experiments depends on the measurement time. To clarify this respect, we have proposed an idealized method measuring the momentum transfer, and investigated the fluctuation measured in this experiment. Our experimental proposal have shown that the prescription proposed by Landau and Lifshitz does not necessarily work in short measurement time. It is interesting to experimentally detect the difference between the fluctuation of the mechanical pressure and that of the microcanonical pressure.

\begin{acknowledgements}
The authors thank Yoshi Oono for his useful comments. The present work was supported by KAKENHI Nos. 25103002 and 17H01148.
\end{acknowledgements}



\end{document}